# Holographic image reconstruction with phase recovery and autofocusing using recurrent neural networks

**LUZHE HUANG**[1,2,3], **TAIRAN LIU**[1,2,3], **XILIN YANG**[1,2,3], **YI LUO**[1,2,3], **YAIR RIVENSON**[1,2,3], **AYDOGAN OZCAN**[1,2,3,4,*]

[1] Electrical and Computer Engineering Department, University of California, Los Angeles, California 90095, USA
[2] Bioengineering Department, University of California, Los Angeles, California 90095, USA
[3] California Nano Systems Institute (CNSI), University of California, Los Angeles, California 90095, USA
[4] David Geffen School of Medicine, University of California, Los Angeles, California 90095, USA
* Corresponding author: ozcan@ucla.edu

**Abstract**

Digital holography is one of the most widely used label-free microscopy techniques in biomedical imaging. Recovery of the missing phase information of a hologram is an important step in holographic image reconstruction. Here we demonstrate a convolutional recurrent neural network (RNN) based phase recovery approach that uses multiple holograms, captured at different sample-to-sensor distances to rapidly reconstruct the phase and amplitude information of a sample, while also performing autofocusing through the same network. We demonstrated the success of this deep learning-enabled holography method by imaging microscopic features of human tissue samples and Papanicolaou (Pap) smears. These results constitute the first demonstration of the use of recurrent neural networks for holographic imaging and phase recovery, and compared with existing methods, the presented approach improves the reconstructed image quality, while also increasing the depth-of-field and inference speed.

# 1. Introduction

Holography provides a powerful tool to image biological samples, with minimal sample preparation, i.e., without the need for staining, fixation or labeling[1–15]. The past decades have seen impressive progress in digital holography field, especially in terms of image reconstruction and quantitative phase imaging (QPI) methods, also providing some unique advantages over traditional microscopic imaging modalities by demonstrating field-portable and cost-effective microscopes for high-throughput imaging, biomedical and sensing applications, among others[1,11,16–32]. One core element in all of these holographic imaging systems is the phase recovery step, since an opto-electronic sensor array only records the intensity of the electromagnetic field impinging on the sensor plane. To retrieve the missing phase information of a sample, a wide range of phase retrieval algorithms have been developed[33–46]; some of these existing algorithms follow a physical model of wave propagation and involve multiple iterations, typically between the hologram and the object planes, in order to recover the missing phase information[34,35,37,47–49]. Recently, deep learning-based phase retrieval algorithms have also been demonstrated to reconstruct a hologram using a trained neural network[39–46,50,51]. These deep learning-based algorithms present advantages over conventional iterative phase recovery methods by creating speckle- and twin-image artifact-free object reconstructions in a single-pass forward through a neural network (i.e., without iterations) and provide additional improvements such as faster image reconstruction and extended depth-of-field (DOF)[41], also enabling cross-modality image transformations, for example matching the color and spatial contrast of brightfield microscopy in the reconstructed hologram[44].

Here we introduce a new deep learning-based holographic image reconstruction and phase retrieval algorithm that is based on a convolutional recurrent neural network (RNN), trained using a generative adversarial network (GAN). As shown in Fig. 1(a), this recurrent holographic (RH) imaging framework uses multiple (M) input holograms that are back-propagated using zero-phase onto a common axial plane to simultaneously perform autofocusing and phase retrieval at its output inference. The efficacy of this method, which we term RH-M, was demonstrated by holographic imaging of human lung tissue sections. Furthermore, by enhancing RH-M with a dilated (D) convolution kernel (Fig. 1(b)), we demonstrate the same autofocusing and phase retrieval performance without the need for any free-space back-propagation (FSP) step, i.e., the acquired raw holograms of an object are directly used as inputs to a trained RNN for in-focus image reconstruction at its output. This second method, termed RH-MD, is more suitable for relatively sparse samples and its success was demonstrated by holographic imaging of Papanicolaou (Pap) smear samples.

When compared with existing phase retrieval and holographic image reconstruction algorithms, RH-M and RH-MD framework introduces important advantages including superior reconstruction quality and speed, as well as extended DOF through its autofocusing feature. As an example, for imaging lung tissue sections, RH-M achieved ~40% quality improvement over existing deep learning-based holographic reconstruction methods in terms of the amplitude root mean squared error (RMSE), and was ~15-fold faster in its inference speed compared to iterative phase retrieval algorithms using the same input holograms. Our results establish the first demonstration of the use of RNNs in holographic imaging and phase recovery, and the presented framework would be broadly useful for various coherent imaging modalities.

## 2. Results

### 2.1 Holographic imaging using RH-M

To demonstrate the efficacy of RH-M for phase recovery and autofocusing, we trained and tested the RNN (Fig. 1(b)) using human lung tissue sections, imaged with a lensfree in-line holographic microscope[22,23,38,52] (see the Materials and Methods section). We used three training slides, covering ~60 mm² unique tissue sample field-of-view and one testing slide, covering ~20 mm² tissue field-of-view; all of these tissue samples were taken from different patients. In the training phase, RH-M randomly took $M$=2 input holograms with random sample-to-sensor distances ranging from 350 μm to 550 μm, i.e., constituting an axial training range of $450 \pm 100$ μm; each one of these randomly selected holograms was then propagated to $\bar{z}_2 = 450$ μm, without any phase recovery step, i.e., using zero phase. The real and imaginary parts of the resulting complex fields were used as training inputs to RH-M model, where the corresponding ground truth complex images of the same samples were obtained using an iterative multi-height phase retrieval (MH-PR) algorithm[14,28] that processed 8 holograms acquired at different sample-to-sensor distances; see the Materials and Methods section for further details. In the blind testing phase, to demonstrate the success of the trained RNN model, $M$=2 holograms of the testing slide (from a different patient, not used during the training) were captured at sample-to-sensor distances of 423.7 μm and 469.7 μm. Both of these test holograms were also back-propagated (using zero phase) to $\bar{z}_2 = 450$ μm, which formed the complex fields $\mathbf{U}_i$; the real and imaginary parts of $\mathbf{U}_i$ were used as inputs to RH-M to test its inference. The results of the RH-M blind inference with these inputs are summarized in Figure 2, which reveals that the output of the RNN, i.e., $\hat{\mathbf{U}}$, very well matches the ground truth $\mathbf{U}_{GT}$ obtained through the iterative MH-PR algorithm that used 8 in-line holograms with accurate knowledge of the sample-to-sensor distances of each hologram plane.

To further analyze RH-M inference performance, we performed a study by feeding the trained RNN with $M$=2 input holograms, captured at various different *combinations* of defocusing distances, i.e., $\Delta z_{2,1}$, and $\Delta z_{2,2}$; the results of this analysis are summarized in Figure 3. First, these results reveal that the presented framework is successful even when $\Delta z_{2,1} = \Delta z_{2,2}$, which corresponds to the case where the two input holograms are identical. Second, the best image reconstruction performance, in terms of the amplitude channel structural similarity index (SSIM) is achieved with a defocus combination of $\Delta z_{2,1} = 3.9$ μm, and $\Delta z_{2,2} = -11.6$ μm (see the red rectangle in Fig. 3(a)), indicating that an axial distance between the two input holograms is preferred to yield better inference for RH-M. The SSIM results reported in Figure 3(b) further illustrate that RH-M can consistently recover the complex object information with various different $\Delta z_{2,1}$ and $\Delta z_{2,2}$ combinations, ranging from -67.0 μm to 35.5 μm, i.e., spanning an axial defocus distance of >100 μm. Figure 3(c) also reports the amplitude and phase RMSE values of the retrieved complex fields with respect to the ground truth field, confirming that RH-M covers an extended DOF of at least 100 μm, matching its axial training range. Furthermore, consistent with the visual reconstruction results reported in Fig. 3(a), for the case of $\Delta z_{2,1} = \Delta z_{2,2}$, RH-M can successfully recover the object fields, but with relatively degraded SSIM values, as indicated by the diagonal entries in Fig. 3(b).

### 2.2 Holographic imaging using RH-MD

RH-M framework can also be extended to perform phase recovery and autofocusing directly from input holograms, without the need for free-space backpropagation using zero-phase and a rough estimate of the sample-to-sensor distance, $\bar{z}_2$ (see Fig. 1(a)). For this goal, we enhanced the RH-M framework by replacing the standard convolutional layers with dilated convolutional layers[53] as shown in Fig. 1(b); we refer to this special case as RH-MD. This change enlarged the receptive field of the network[54], which provides RH-MD the capability to process diffraction patterns over a relatively larger area without increasing the number of trainable parameters, while also opening up the possibility to directly perform phase recovery and autofocusing from raw input holograms. To demonstrate this capability, we trained and tested the RH-MD framework on Pap smear samples imaged by the same lensfree holographic microscopy platform. Here, the training dataset contains raw in-line holograms with random sample-to-sensor distances ranging from 400 μm to 600 μm, i.e., constituting the training range of $500 \pm 100$ μm; testing image dataset contains raw holograms of sample fields-of-view that were never seen by the network before. Figure 4 summarizes the blind inference results of RH-MD and its performance comparison against the results of RH-M for the same test regions of interest. In Fig. 4(a), the outputs of both RH-M and RH-MD very well match the ground truth complex fields – however, different from RH-M, RH-MD used the raw input holograms without the need for any free-space backpropagation over $\bar{z}_2$. As shown in the region-of-interest (ROI) labeled by the green box in Fig. 4(a,b), both RH-M and RH-MD are able to suppress the concentric ring artifacts induced by some out-of-focus particles (pointed by the yellow arrows); such particles lie outside of the sample plane and therefore, are treated as interference artifacts, and removed by both RH-M and RH-MD since they were trained using two dimensional (2D) samples. Similar observations were also made in earlier deep learning-based holographic image reconstruction methods[39].

Figure 4(c) further illustrates a comparison of the amplitude and phase SSIM values of the output images of RH-M and RH-MD, with respect to the ground truth field. Without the need for back-propagation of input holograms, RH-MD appears to be more robust compared to RH-M when the input holograms are obtained at the same height ($\Delta z_{2,1} = \Delta z_{2,2}$), as indicated by the diagonal entries in Fig. 4(c). These advantages of RH-MD are observed for microscopic imaging of relatively sparse samples such as Pap smear slides reported here; however, for connected tissue sections such as histopathology samples shown in Figs. 2-3, we observed that the RH-MD inference performance significantly degraded compared to RH-M since for such connected and spatially-dense samples the differential advantage brought by dilated convolutional layers disappears due to cross-talk from other features/samples within the field-of-view. Therefore, unlike RH-M, sample sparsity is found to be a requirement for blind phase retrieval, holographic image reconstruction and autofocusing using RH-MD.

## 3. Discussion

The presented work reports the first demonstration of an RNN-based phase retrieval framework that incorporates sequential input holograms to perform holographic image reconstruction with autofocusing. To better highlight the advantages of this RNN-based framework over existing neural network-based phase recovery and holographic image reconstruction methods, we compared the performance of RH-M to an earlier method, termed Holographic Imaging using Deep Learning for Extended Focus (HIDEF)[41]. This previous framework, HIDEF, used a trained convolutional neural network (CNN) to perform both autofocusing and phase retrieval with a single hologram that is back-propagated using zero-phase. For providing a quantitative comparison of RH-M and HIDEF, we tested both algorithms using 3 holograms of lung tissue sections that were acquired at different sample-to-sensor distances of 383, 438.4 and 485.5 μm (see Fig. 5). Both HIDEF and RH-M used the same $\bar{z}_2 = 450$ μm in the initial back-propagation step (using zero-phase), and by design HIDEF takes in one hologram each time. As shown in Figs. 5(a-c), RH-M achieved a superior image reconstruction quality compared to HIDEF through the utilization of multiple holograms, providing ~40% improvement in terms of the amplitude RMSE. Figure 5(e) further summarizes the mean and the standard deviation of RMSE values for RH-M and HIDEF output images, showing the statistical significance of this improvement. Benefitting from its recurrent scheme, RH-M can easily adapt to a variable number of input holograms, e.g., $M=2$ or $M=3$, although the RNN was trained with fixed number of input holograms. By adding one more input hologram, the RH-M output is further improved as illustrated in Fig. 5(e); see the results for $M=2$ vs. $M=3$. Furthermore, it is important to note that RH-M has a very good inference stability with respect to the order of the input holograms. Figure 5 illustrates the consistency of the retrieved field by RH-M over different selections and/or permutations of the input holograms. This feature of RH-M provides great flexibility and advantage in the acquisition of raw holograms without the need for accurate axial sampling or a fixed scanning direction/grid.

Another important advantage of RH-M is the extended DOF that it offers, over both HIDEF and MH-PR results. In Fig. 6, simulated input holograms were generated with sample-to-sensor distances ranging from 300 μm to 600 μm, and then back-propagated using zero-phase onto the same axial plane, $\bar{z}_2 = 450$ μm. Fig. 6(a) illustrates the RMSE values of the amplitude channel of each output field as a function of the defocus distance $\Delta z_2$ generated by RH-M (red lines and pink shadow), MH-PR (grey lines and shadow) and HIDEF (blue line), where RH-M ($M=2$) and MH-PR ($M=2$) used the average defocus distance of the two input holograms. For RH-M and MH-PR, where two input holograms are present, the mean and optimal RMSE values over all $\Delta z_{2,1}$, $\Delta z_{2,2}$ defocus combinations are shown by the solid and dashed lines, respectively, and the pink and grey shadows indicate the standard deviations of the RMSE values for each case. The quantitative comparisons presented in Fig. 6(a) clearly reveal the DOF advantage of RH-M output images, also demonstrating better image quality as evidenced by the lower RMSE values, which is also supported by the visual comparisons provided in Fig. 6(b).

Finally, in Table 1, we compared the output inference (or image reconstruction) speed of RH-M, RH-MD, HIDEF and MH-PR algorithms using Pap smear and lung tissue samples. As shown in Table 1, among these phase retrieval and holographic image reconstruction algorithms, RH-M and RH-MD are the fastest, achieving ~50-fold and ~15-fold reconstruction speed improvement compared with MH-PR ($M=8$ and $M=2$, respectively); unlike RH-M or RH-MD, the performance of MH-PR is also dependent on the accuracy of the knowledge/measurement of the sample-to-sensor distance for each raw hologram, which is not the case for the RNN-based hologram reconstruction methods reported here.

In summary, our results present advantages over existing phase retrieval and holographic image reconstruction methods in terms of image quality, DOF and inference speed. This RNN-based image reconstruction method opens up new opportunities for a variety of coherent microscopy modalities and related applications.

## 4. Materials and Methods

*4.1 Holographic microscopy hardware and imaging of samples*

Raw holograms were collected using a lensfree in-line holographic microscopy setup shown in Fig. 1(a). A broadband light source (WhiteLase Micro, NKT Photonics) filtered by an acousto-optic tunable filter (AOTF) was used as the illumination source (530 nm). A complementary metal-oxide semiconductor (CMOS) color image sensor (IMX 081, Sony, pixel size of 1.12 μm) was used to capture the raw holograms. The sample was directly placed between the illumination source and the sensor plane with a sample-to-source distance ($z_1$) of ~5-10 cm, and a sample-to-sensor distance ($z_2$) of ~300-600 μm. The image sensor was attached to a 3D positioning stage (MAX606, Thorlabs, Inc.) to capture holograms at different lateral and axial positions to perform pixel super-resolution[22,55,56] and multi-height phase recovery[22,23,38], respectively. All imaging hardware was controlled by a customized LabVIEW program to complete the data acquisition automatically.

All the human samples imaged in this work were obtained after deidentification of the patient information and were prepared from existing specimen; therefore, this work did not interfere with standard practices of medical care or sample collection procedures.

### 4.2 Pixel super-resolution

A pixel super-resolution algorithm was implemented to enhance the hologram resolution and bring the effective image pixel size from 2.24 μm down to 0.37 μm. To perform this, in-line holograms at 6-by-6 lateral positions were captured with sub-pixel spacing using a 3D positioning stage (MAX606, Thorlabs, Inc.). The accurate relative displacements/shifts were estimated by an image correlation-based algorithm and the high-resolution hologram was generated using the shift-and-add algorithm[55,57]. The resulting super-resolved holograms (also referred to as raw holograms) were used for phase retrieval and holographic imaging, as reported in the Results section.

### 4.3 Angular spectrum-based field propagation

The angular spectrum-based field propagation[58] was employed as an essential building block for both the holographic autofocusing and the multi-height phase recovery algorithms. This numerical propagation procedure enables one to propagate the initial complex optical field at $z = z_0$ to obtain the complex optical field at $z = z_0 + \Delta z$. A 2D Fourier transform is first applied on the initial complex optical field $U(x, y; z_0)$ and the resulting angular spectrum is then multiplied by a spatial frequency-dependent phase factor parametrized by the wavelength, refractive index of the medium, and the propagation distance in free-space ($\Delta z$). Finally, to retrieve the complex optical field at $z = z_0 + \Delta z$, i.e., $U(x, y; z_0 + \Delta z)$, an inverse 2D Fourier transform is applied.

### 4.4 Multi-height phase recovery

In-line holograms at different axial positions, with e.g., ~15 μm spacing, were captured to perform MH-PR[22]. The relative axial distances between different holograms were estimated using an autofocusing algorithm[59] based on the edge sparsity criterion. The iterative MH-PR algorithm first takes the amplitude of the hologram captured at the first height (i.e., $z_{2,1}$) and pads an all-zero phase channel to it. It then propagates the resulting field to different hologram heights, where the amplitude channel is updated at each height by averaging the amplitude channel of the propagated field with the measured amplitude of the hologram acquired at that corresponding height [60]. This iterative algorithm converges typically after 10-30 iterations, where one iteration is complete after all the measured holograms have been used as part of the multi-height amplitude updates. Finally, the converged complex field is backpropagated onto the sample plane using the sample-to-sensor distance determined by the autofocusing algorithm. To generate the ground truth images for the network training and testing phases, in-line holograms at 8 different heights were used for both the lung and the Pap smear samples reported in this work.

### 4.5 Network structure

RH-M and RH-MD adapt the GAN framework for their training, which is depicted in Fig. 7. As shown in Fig. 1(b), RH-M and RH-MD, i.e., the generators, share the same convolutional

RNN structure[61], which consists of down- and up-sampling paths with consecutive convolutional blocks at 4 different scales. Between the convolutional blocks at the same scale on the down- and up-sampling paths, a convolutional recurrent block connects them and passes high frequency features. The number of channels of the i-th convolution layer at the k-th convolution block (where $k$ = 1, 2, 3, 4, and $i$ = 1, 2) of RH-M and RH-MD are $20\times 2^{k-3+i}$ and $16\times 2^{k-3+i}$ respectively. For RH-MD, the convolution layer in each block applies a dilated kernel with a dilation rate of 2 (Fig. 1(b)). The convolutional recurrent block follows the structure of one convolutional gated recurrent unit (CGRU) layer and one $1\times 1$ convolution layer as in Ref. 61. As illustrated in Fig. 1(c), a standard CNN with 5 convolutional blocks and 2 dense layers was adapted to serve as the discriminator in the GAN framework. The k-th convolutional block of the discriminator has two convolutional layers with $20\times 2^{k-1}$ channels and each layer uses a 3×3 kernel with a stride of 1.

## 4.6 Network implementation

RH-M and RH-MD were implemented using TensorFlow with Python and CUDA environments, and trained on a computer with Intel Xeon W-2195 processor, 256 GB memory and one NVIDIA RTX 2080 Ti graphic processing unit (GPU). In the training phase, for each field-of-view, $M_{train}$ holograms were randomly selected from different heights (sample-to-sensor distances) as the network input, and then the corresponding output field of RH-M or RH-MD was sent to the discriminator ($D$) network.

The generator loss $L_G$ is the weighted sum of three different loss terms: (1) pixel-wise mean absolute error (MAE), $L_{MAE}$, (2) multi-scale structural similarity (MSSSIM) $L_{MSSSIM}$ between the network output $\hat{y}$ and the ground truth $y$, and (3) the adversarial loss $L_{G,D}$ from the discriminator network. Based on these, the total generator loss can be expressed as:

$$L_G = \alpha L_{MAE} + \beta L_{MSSSIM} + \gamma L_{G,D} \tag{1}$$

where $\alpha, \beta, \gamma$ are relative weights, empirically set as 3, 1, 0.5, respectively. The MAE and MSSSIM losses are defined as:

$$L_{MAE} = \frac{\sum_{i=1}^{n}|\hat{y}(i) - y(i)|}{n} \tag{2}$$

$$L_{MSSSIM} = 1 - \left[\frac{2\mu_{\hat{y}_m}\mu_{y_m} + C_1}{\mu_{\hat{y}_m}^2 + \mu_{y_m}^2 + C_1}\right]^{\alpha_m} \cdot \prod_{j=1}^{m}\left[\frac{2\sigma_{\hat{y}_j}\sigma_{y_j} + C_2}{\sigma_{\hat{y}_j}^2 + \sigma_{y_j}^2 + C_2}\right]^{\beta_j}\left[\frac{\sigma_{\hat{y}_j y_j} + C_3}{\sigma_{\hat{y}_j}\sigma_{y_j} + C_3}\right]^{\gamma_j} \tag{3}$$

where $n$ is the total number of pixels in $y$, and $\hat{y}_j, y_j$ are $2^{j-1}$ downsampled images of $\hat{y}, y$, respectively. $\mu_y, \sigma_y^2$ represent the mean and variance of the image $y$, respectively, while $\sigma_{\hat{y}_j y_j}$ is the covariance between $\hat{y}_j$ and $y_j$. $C_1, C_2, C_3, \alpha_m, \beta_j, \gamma_j, m$ are pre-defined empirical hyperparameters[62]. The adversarial loss $L_{G,D}$ and the total discriminator loss $L_D$ are calculated as follows:

$$L_{G,D} = [D(\hat{y}) - 1]^2 \tag{4}$$

$$L_D = \frac{1}{2}D(\hat{y})^2 + \frac{1}{2}[D(y) - 1]^2 \tag{5}$$

Adam optimizers[63] with decaying learning rates, initially set as $5\times 10^{-5}$ and $1\times 10^{-6}$, were employed for the optimization of the generator and discriminator networks, respectively. After ~30 hours of training, corresponding to ~10 epochs, the training was stopped to avoid possible overfitting[64].

In the testing phase of RH-M and RH-MD, the convolutional RNN was optimized for mixed precision computation. In general, a trained RNN can be fed with input sequences of variable length. In our experiments, RH-M/RH-MD was trained on datasets with fixed number of inputs (holograms) to save time, i.e., fixed $M_{train}$, and later tested on testing data with no more than $M_{train}$ input holograms (i.e., $M_{test} \leqslant M_{train}$). In consideration of the convergence of recurrent units, shorter testing sequences (where $M_{test} < M_{train}$) were replication-padded to match the length of the training sequences $M_{train}$. For example, in Fig. 5, the RH-M was trained solely on datasets with 3 input holograms and tested with 2 or 3 input holograms.

HIDEF networks were trained in the same way as detailed in Ref. 41. Blind testing and comparison of all the algorithms (HIDEF, RH-M, RH-MD and MH-PR) were implemented on a computer with Intel Core i9-9820X processor, 128 GB memory and one NVIDIA TITAN RTX graphic card using GPU acceleration, and the details, including the number of parameters and inference times are summarized in Table 1.

## References

1   Seo S, Su T-W, Tseng DK, Erlinger A, Ozcan A. Lensfree holographic imaging for on-chip cytometry and diagnostics. *Lab Chip* 2009; **9**: 777–787.
2   Hsieh C-L, Grange R, Pu Y, Psaltis D. Three-dimensional harmonic holographic microcopy using nanoparticles as probes for cell imaging. 2009; : 12.
3   Shaked NT, Rinehart MT, Wax A. Dual-interference-channel quantitative-phase microscopy of live cell dynamics. *Opt Lett* 2009; **34**: 767.
4   Kou SS, Waller L, Barbastathis G, Sheppard CJR. Transport-of-intensity approach to differential interference contrast (TI-DIC) microscopy for quantitative phase imaging. *Opt Lett* 2010; **35**: 447.
5   Popescu G. *Quantitative phase imaging of cells and tissues*. McGraw-Hill: New York, 2011.
6   Mir M, Bhaduri B, Wang R, Zhu R, Popescu G. Quantitative Phase Imaging. In: *Progress in Optics*. Elsevier, 2012, pp 133–217.
7   Shaked NT. Quantitative phase microscopy of biological samples using a portable interferometer. *Opt Lett* 2012; **37**: 2016.
8   Su T-W, Xue L, Ozcan A. High-throughput lensfree 3D tracking of human sperms reveals rare statistics of helical trajectories. *Proceedings of the National Academy of Sciences* 2012; **109**: 16018–16022.
9   Chhaniwal V, Singh ASG, Leitgeb RA, Javidi B, Anand A. Quantitative phase-contrast imaging with compact digital holographic microscope employing Lloyd's mirror. *Opt Lett* 2012; **37**: 5127.
10  Jericho MH, Kreuzer HJ, Kanka M, Riesenberg R. Quantitative phase and refractive index measurements with point-source digital in-line holographic microscopy. *Appl Opt* 2012; **51**: 1503.
11  Greenbaum A, Zhang Y, Feizi A, Chung P-L, Luo W, Kandukuri SR *et al.* Wide-field computational imaging of pathology slides using lens-free on-chip microscopy. *Science Translational Medicine* 2014; **6**: 267ra175.
12  Tian L, Waller L. Quantitative differential phase contrast imaging in an LED array microscope. *Opt Express* 2015; **23**: 11394.
13  Merola F, Memmolo P, Miccio L, Savoia R, Mugnano M, Fontana A *et al.* Tomographic flow cytometry by digital holography. *Light Sci Appl* 2017; **6**: e16241–e16241.
14  Park Y, Depeursinge C, Popescu G. Quantitative phase imaging in biomedicine. *Nature Photon* 2018; **12**: 578–589.
15  Barbastathis G, Ozcan A, Situ G. On the use of deep learning for computational imaging. *Optica* 2019; **6**: 921.

16  Ferraro P, Grilli S, Alfieri D, De Nicola S, Finizio A, Pierattini G *et al.* Extended focused image in microscopy by digital holography. *Opt Express* 2005; **13**: 6738.
17  Ozcan A, Demirci U. Ultra wide-field lens-free monitoring of cells on-chip. *Lab Chip* 2008; **8**: 98–106.
18  Tseng D, Mudanyali O, Oztoprak C, Isikman SO, Sencan I, Yaglidere O *et al.* Lensfree microscopy on a cellphone. *Lab Chip* 2010; **10**: 1787.
19  Mudanyali O, Tseng D, Oh C, Isikman SO, Sencan I, Bishara W *et al.* Compact, light-weight and cost-effective microscope based on lensless incoherent holography for telemedicine applications. *Lab Chip* 2010; **10**: 1417.
20  Shi K, Li H, Xu Q, Psaltis D, Liu Z. Coherent Anti-Stokes Raman Holography for Chemically Selective Single-Shot Nonscanning 3D Imaging. *Phys Rev Lett* 2010; **104**: 093902.
21  Yuan C, Situ G, Pedrini G, Ma J, Osten W. Resolution improvement in digital holography by angular and polarization multiplexing. *Appl Opt* 2011; **50**: B6.
22  Greenbaum A, Ozcan A. Maskless imaging of dense samples using pixel super-resolution based multi-height lensfree on-chip microscopy. *Opt Express* 2012; **20**: 3129.
23  Greenbaum A, Luo W, Su T-W, Göröcs Z, Xue L, Isikman SO *et al.* Imaging without lenses: achievements and remaining challenges of wide-field on-chip microscopy. *Nat Methods* 2012; **9**: 889–895.
24  Paturzo M, Finizio A, Memmolo P, Puglisi R, Balduzzi D, Galli A *et al.* Microscopy imaging and quantitative phase contrast mapping in turbid microfluidic channels by digital holography. *Lab Chip* 2012; **12**: 3073.
25  Bao P, Situ G, Pedrini G, Osten W. Lensless phase microscopy using phase retrieval with multiple illumination wavelengths. *Appl Opt* 2012; **51**: 5486.
26  Pham HV, Bhaduri B, Tangella K, Best-Popescu C, Popescu G. Real Time Blood Testing Using Quantitative Phase Imaging. *PLoS ONE* 2013; **8**: e55676.
27  M. Rotermund L, Samson J, Kreuzer HJ. A Submersible Holographic Microscope for 4-D In-Situ Studies of Micro-Organisms in the Ocean with Intensity and Quantitative Phase Imaging. *J Marine Sci Res Dev* 2015; **06**. doi:10.4172/2155-9910.1000181.
28  Ozcan A, McLeod E. Lensless Imaging and Sensing. *Annu Rev Biomed Eng* 2016; **18**: 77–102.
29  Majeed H, Sridharan S, Mir M, Ma L, Min E, Jung W *et al.* Quantitative phase imaging for medical diagnosis. *J Biophoton* 2017; **10**: 177–205.
30  Rawat S, Komatsu S, Markman A, Anand A, Javidi B. Compact and field-portable 3D printed shearing digital holographic microscope for automated cell identification. *Appl Opt* 2017; **56**: D127.
31  Göröcs Z, Tamamitsu M, Bianco V, Wolf P, Roy S, Shindo K *et al.* A deep learning-enabled portable imaging flow cytometer for cost-effective, high-throughput, and label-free analysis of natural water samples. *Light Sci Appl* 2018; **7**: 66.
32  Wu Y, Ozcan A. Lensless digital holographic microscopy and its applications in biomedicine and environmental monitoring. *Methods* 2018; **136**: 4–16.
33  Teague MR. Deterministic phase retrieval: a Green's function solution. *J Opt Soc Am* 1983; **73**: 1434.
34  Allen LJ, Oxley MP. Phase retrieval from series of images obtained by defocus variation. *Optics Communications* 2001; **199**: 65–75.
35  Marchesini S. Invited Article: A unified evaluation of iterative projection algorithms for phase retrieval. *Review of Scientific Instruments* 2007; **78**: 011301.
36  Luo W, Greenbaum A, Zhang Y, Ozcan A. Synthetic aperture-based on-chip microscopy. *Light Sci Appl* 2015; **4**: e261–e261.
37  Luo W, Zhang Y, Göröcs Z, Feizi A, Ozcan A. Propagation phasor approach for holographic image reconstruction. *Sci Rep* 2016; **6**: 22738.
38  Rivenson Y, Wu Y, Wang H, Zhang Y, Feizi A, Ozcan A. Sparsity-based multi-height phase recovery in holographic microscopy. *Sci Rep* 2016; **6**: 37862.

39  Rivenson Y, Zhang Y, Günaydın H, Teng D, Ozcan A. Phase recovery and holographic image reconstruction using deep learning in neural networks. *Light Sci Appl* 2018; **7**: 17141.
40  Goy A, Arthur K, Li S, Barbastathis G. Low Photon Count Phase Retrieval Using Deep Learning. *Phys Rev Lett* 2018; **121**: 243902.
41  Wu Y, Rivenson Y, Zhang Y, Wei Z, Günaydin H, Lin X *et al.* Extended depth-of-field in holographic imaging using deep-learning-based autofocusing and phase recovery. *Optica* 2018; **5**: 704.
42  Wang H, Lyu M, Situ G. eHoloNet: a learning-based end-to-end approach for in-line digital holographic reconstruction. *Opt Express* 2018; **26**: 22603.
43  Wang K, Dou J, Kemao Q, Di J, Zhao J. Y-Net: a one-to-two deep learning framework for digital holographic reconstruction. *Opt Lett* 2019; **44**: 4765.
44  Wu Y, Luo Y, Chaudhari G, Rivenson Y, Calis A, de Haan K *et al.* Bright-field holography: cross-modality deep learning enables snapshot 3D imaging with bright-field contrast using a single hologram. *Light Sci Appl* 2019; **8**: 25.
45  Liu T, de Haan K, Bai B, Rivenson Y, Luo Y, Wang H *et al.* Deep Learning-Based Holographic Polarization Microscopy. *ACS Photonics* 2020; **7**: 3023–3034.
46  Deng M, Li S, Goy A, Kang I, Barbastathis G. Learning to synthesize: robust phase retrieval at low photon counts. *Light Sci Appl* 2020; **9**: 36.
47  Yang G, Dong B, Gu B, Zhuang J, Ersoy OK. Gerchberg–Saxton and Yang–Gu algorithms for phase retrieval in a nonunitary transform system: a comparison. *Appl Opt* 1994; **33**: 209.
48  Gureyev TE, Nugent KA. Rapid quantitative phase imaging using the transport of intensity equation. *Optics Communications* 1997; **133**: 339–346.
49  Gureyev TE, Pogany A, Paganin DM, Wilkins SW. Linear algorithms for phase retrieval in the Fresnel region. *Optics Communications* 2004; **231**: 53–70.
50  Liu T, Wei Z, Rivenson Y, Haan K, Zhang Y, Wu Y *et al.* Deep learning‐based color holographic microscopy. *J Biophotonics* 2019; **12**. doi:10.1002/jbio.201900107.
51  Rivenson Y, Wu Y, Ozcan A. Deep learning in holography and coherent imaging. *Light Sci Appl* 2019; **8**: 85.
52  Greenbaum A, Sikora U, Ozcan A. Field-portable wide-field microscopy of dense samples using multi-height pixel super-resolution based lensfree imaging. *Lab Chip* 2012; **12**: 1242.
53  Yu F, Koltun V. Multi-Scale Context Aggregation by Dilated Convolutions. *arXiv:151107122 [cs]* 2016.http://arxiv.org/abs/1511.07122 (accessed 26 Jan2021).
54  Luo W, Li Y, Urtasun R, Zemel R. Understanding the Effective Receptive Field in Deep Convolutional Neural Networks. *arXiv:170104128 [cs]* 2017.http://arxiv.org/abs/1701.04128 (accessed 26 Jan2021).
55  Greenbaum A, Feizi A, Akbari N, Ozcan A. Wide-field computational color imaging using pixel super-resolved on-chip microscopy. *Opt Express* 2013; **21**: 12469.
56  Liu T, de Haan K, Rivenson Y, Wei Z, Zeng X, Zhang Y *et al.* Deep learning-based super-resolution in coherent imaging systems. *Sci Rep* 2019; **9**: 3926.
57  Greenbaum A, Ozcan A. Maskless imaging of dense samples using pixel super-resolution based multi-height lensfree on-chip microscopy. *Opt Express, OE* 2012; **20**: 3129–3143.
58  Goodman JW. *Introduction to Fourier optics*. 3rd ed. Roberts & Co: Englewood, Colo, 2005.
59  Zhang Y, Wang H, Wu Y, Tamamitsu M, Ozcan A. Edge sparsity criterion for robust holographic autofocusing. *Opt Lett* 2017; **42**: 3824–3827.
60  Goodman JW. *Introduction to Fourier optics*. Roberts and Company Publishers, 2005.
61  Huang L, Luo Y, Rivenson Y, Ozcan A. Recurrent neural network-based volumetric fluorescence microscopy. *arXiv:201010781 [physics]* 2020.http://arxiv.org/abs/2010.10781 (accessed 2 Dec2020).

62 Wang Z, Simoncelli EP, Bovik AC. Multiscale structural similarity for image quality assessment. In: *The Thrity-Seventh Asilomar Conference on Signals, Systems & Computers, 2003*. IEEE: Pacific Grove, CA, USA, 2003, pp 1398–1402.

63 Kingma DP, Ba J. Adam: A Method for Stochastic Optimization. *arXiv:14126980 [cs]* 2017.http://arxiv.org/abs/1412.6980 (accessed 13 Jul2020).

64 Prechelt L. Early Stopping - But When? In: Orr GB, Müller K-R (eds). *Neural Networks: Tricks of the Trade*. Springer Berlin Heidelberg: Berlin, Heidelberg, 1998, pp 55–69.

# Figures and Tables

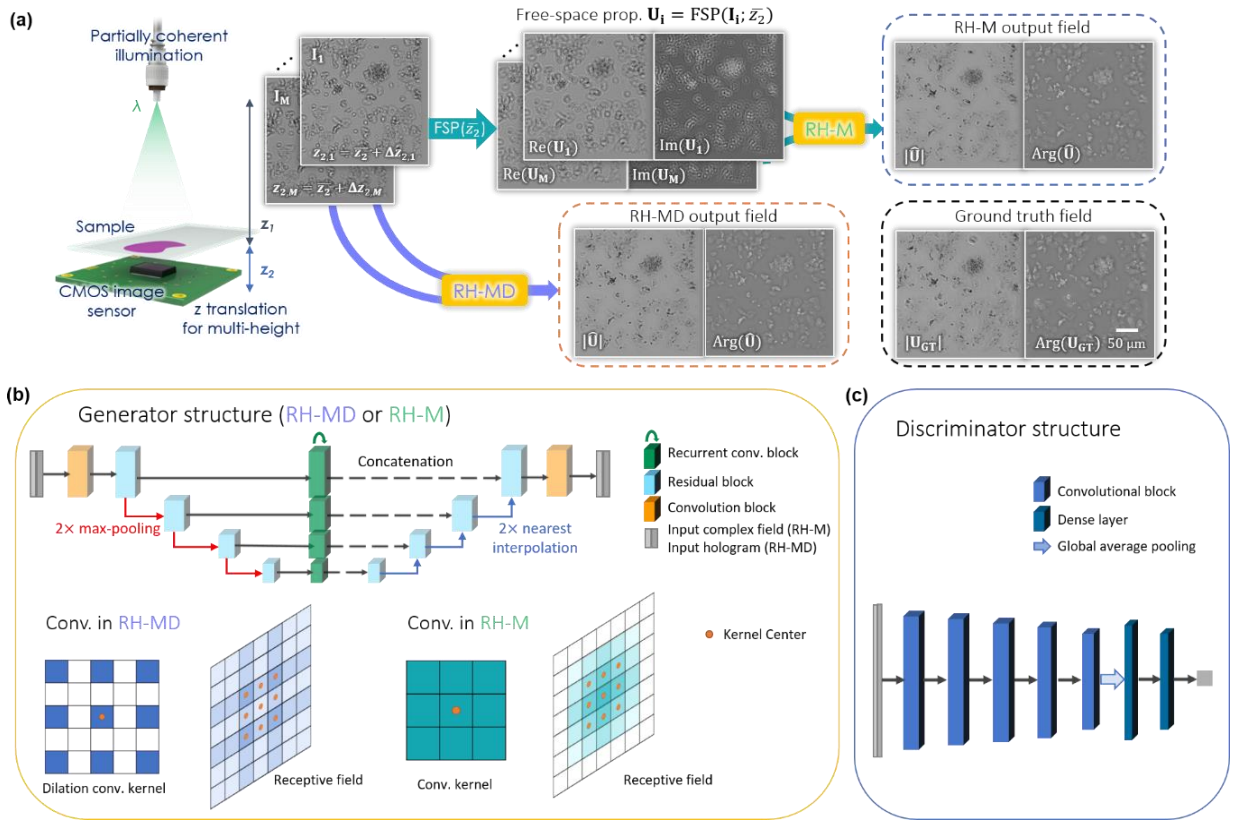

**Fig. 1** Recurrent holographic imaging framework (RH-M and RH-MD). (a) Raw holograms $\mathbf{I}_i$, $i = 1, 2, …, M$, captured at unknown axial positions $z_{2,i}$ located around an arbitrary axial position $\bar{z}_2$. $\mathbf{U}_{GT}$ denotes the ground truth (GT) complex field at the sample plane obtained by iterative multi-height phase retrieval (MH-PR)[22,38] that used 8 holograms acquired at different sample-to-sensor distances. $\hat{\mathbf{U}}$ denotes the retrieved complex field by recurrent holographic imaging networks, and $\mathbf{U}_i$ stands for the complex field resulting from the propagation of $\mathbf{I}_i$ by an axial distance of $\bar{z}_2$ using zero phase (i.e., without any phase recovery step). (b) The generator network structure of RH-M and RH-MD, and examples of the dilated / non-dilated convolutional kernels and the corresponding receptive fields are shown. The input and output images in RH-M / RH-MD have two channels corresponding to the real and imaginary parts of the optical fields, respectively. (c) The discriminator ($D$) network structure used for training of RH-M and RH-MD using a GAN framework. Also see the Materials and Methods section.

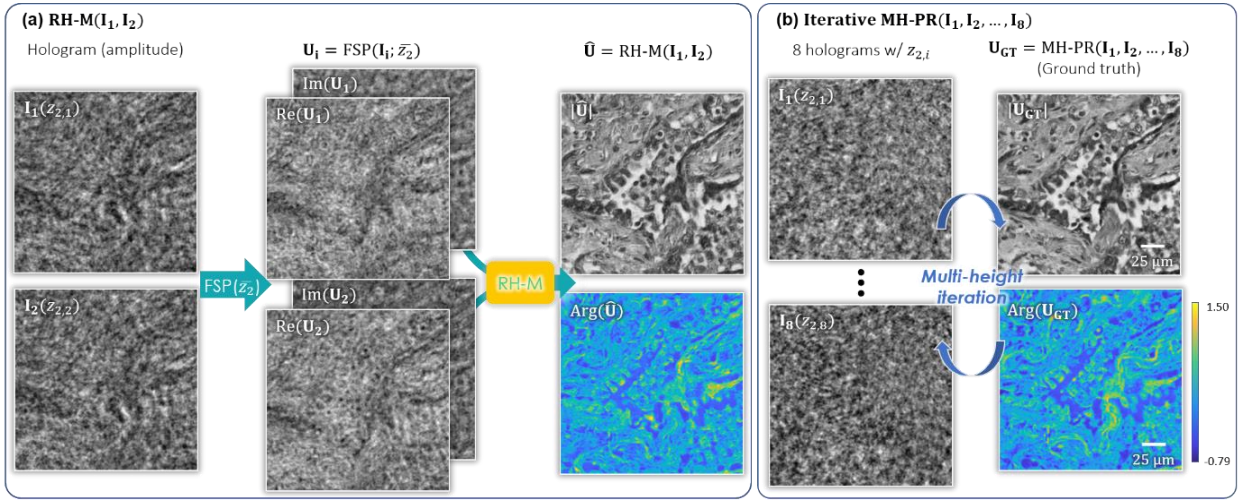

**Fig. 2** Holographic imaging of lung tissue sections. (a) RH-M inference results using $M=2$ input holograms. (b) Holographic imaging with 8 holograms ($I_1 \ldots I_8$) using the iterative MH-PR algorithm, which constitutes our ground truth.

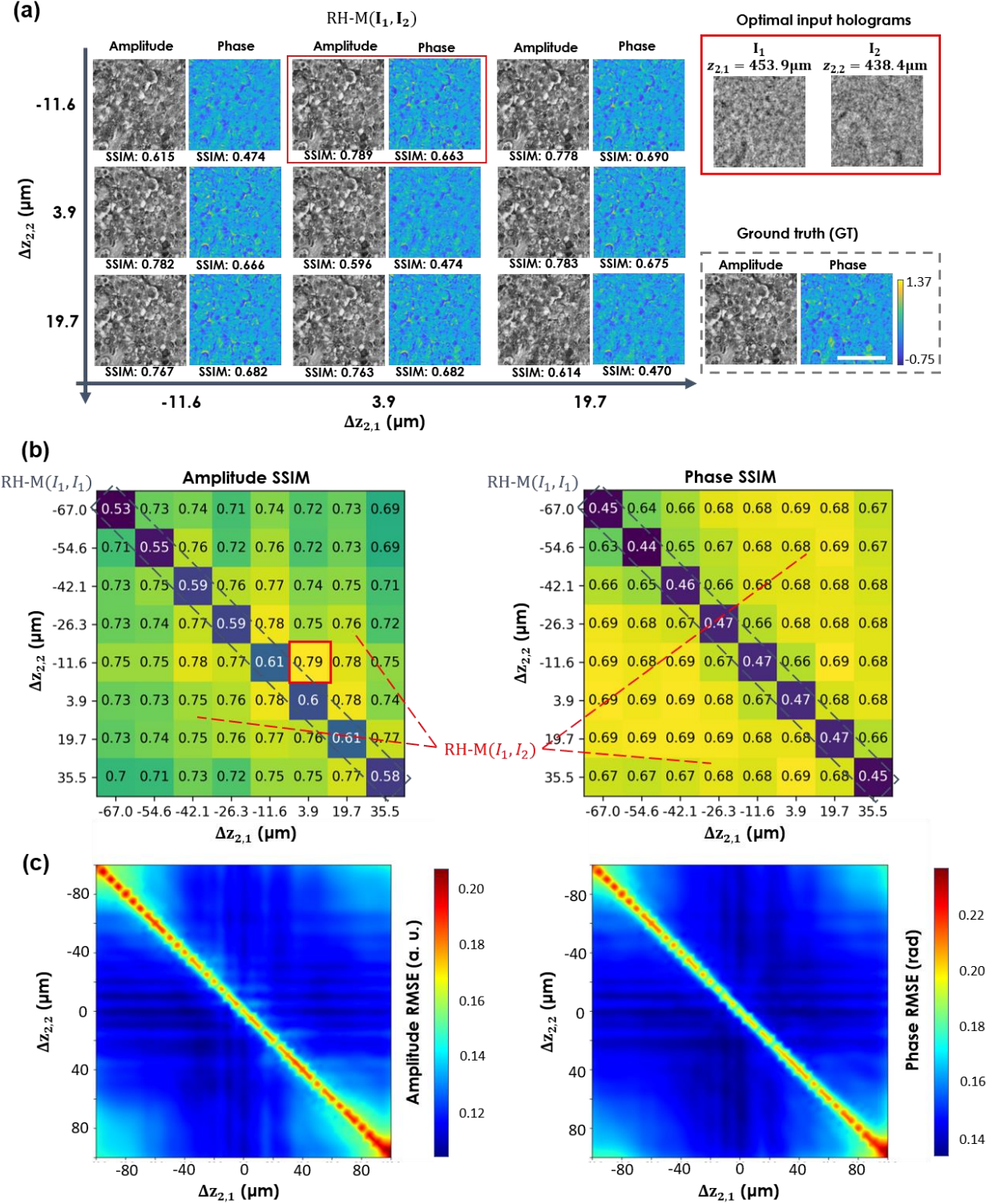

**Fig. 3** Holographic imaging of lung tissue sections using RH-M (*M*=2). (a) The retrieved complex field (RH-M output) using *M*=2 input holograms at various combinations of sample-to-sensor distances. The optimal holographic input combination is highlighted by the red solid line box, corresponding to the RH-M output with the highest amplitude SSIM. The ground truth field obtained by the iterative MH-PR algorithm using 8 holograms/heights is highlighted by the grey dash-line box. (b) Amplitude and phase SSIM values of RH-M output images with *M*=2 input holograms acquired at various heights (sample-to-sensor distances) around $\bar{z}_2 = 450$ μm. (c) Amplitude and phase RMSE values of RH-M output images with *M*=2 input holograms acquired at various heights around $\bar{z}_2 = 450$ μm. Scale bar: 25 μm.

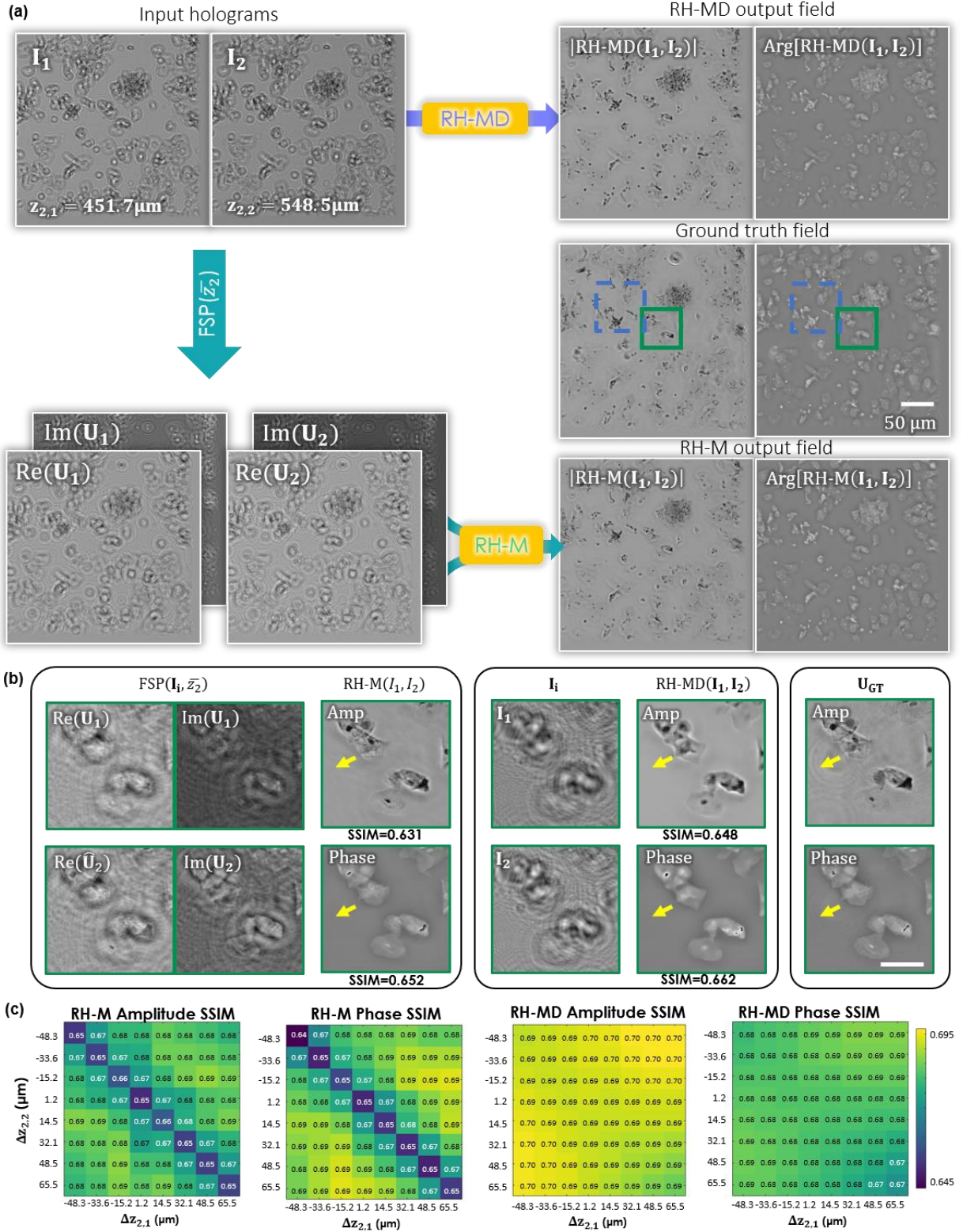

**Fig. 4** Holographic imaging of a Pap smear using RH-M and RH-MD ($M=2$). (a) RH-MD network directly takes in the raw holograms as its input, while RH-M first back-propagates the input holograms using zero phase to $\bar{z}_2$ and then takes in these back-propagated complex fields as its input. The outputs from RH-M and RH-MD both match the ground truth field very well. (b) The expanded regions of interest (ROI) highlighted by green boxes in (a). The yellow arrows point out some concentric ring artifacts in the ground truth image (created by an out-of-plane particle), which are suppressed by both RH-M and RH-MD. Such out-of-focus particles appear in MH-PR image reconstruction results as they are physically expected, whereas they are suppressed in the RNN output images because both RH-M and RH-MD are trained using 2D samples. Scale bar: 10 μm. (c) Amplitude and phase SSIM values of the retrieved field from $M$=2 input holograms at different axial positions around $\bar{z}_2 = 500$ μm. The SSIM values are calculated within the ROI marked by the blue dashed line box in (a).

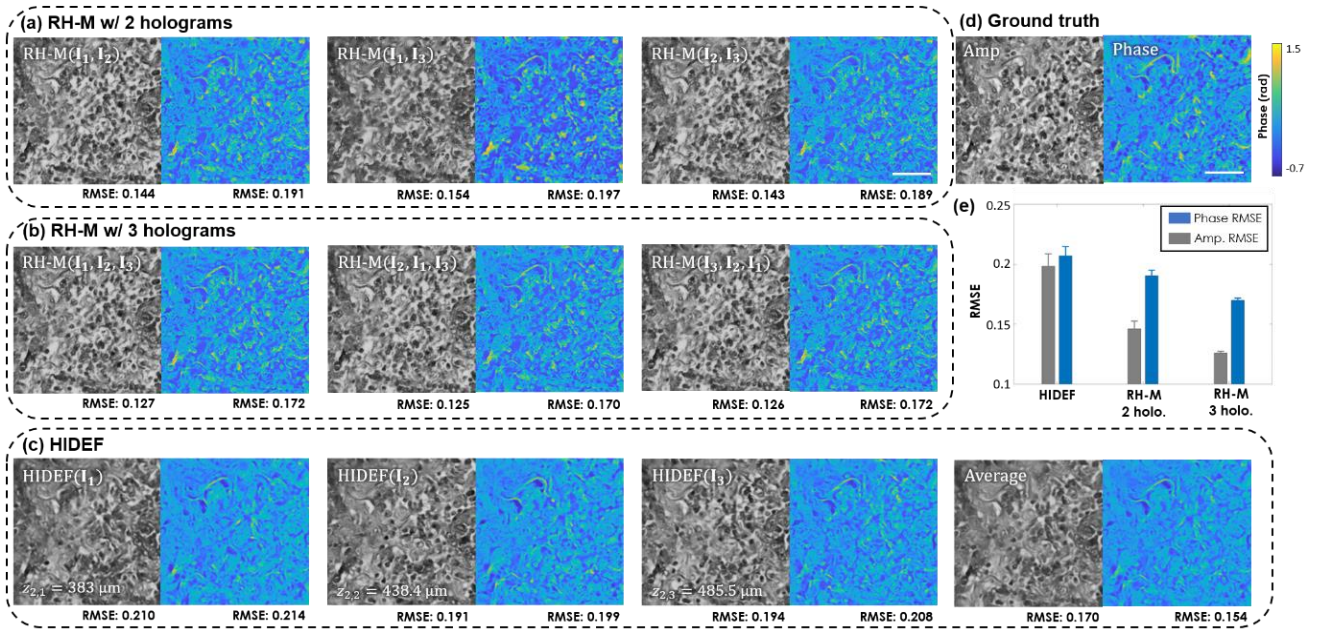

**Fig. 5** RH-M performance comparison against HIDEF using lung tissue sections. (a) The retrieved field by RH-M with $M$=2 input holograms matches the ground truth very well. (b) The retrieved field by RH-M with $M$=3 input holograms provides an improved match against the ground truth. (c) The retrieved field by HIDEF using a single input hologram ($\mathbf{I}_1$, $\mathbf{I}_2$ or $\mathbf{I}_3$). The average field that is reported here is calculated by averaging HIDEF($\mathbf{I}_1$), HIDEF($\mathbf{I}_2$) and HIDEF($\mathbf{I}_3$). (d) Ground truth field obtained by iterative MH-PR that used 8 holograms acquired at different sample-to-sensor distances. (e) The bar plot of the RMSE values of the retrieved field from HIDEF, RH-M with $M$=2 input holograms (6 selections), RH-M with $M$=3 input holograms (6 permutations), with respect to the ground truth field shown in (d). Scale bar: 50 μm.

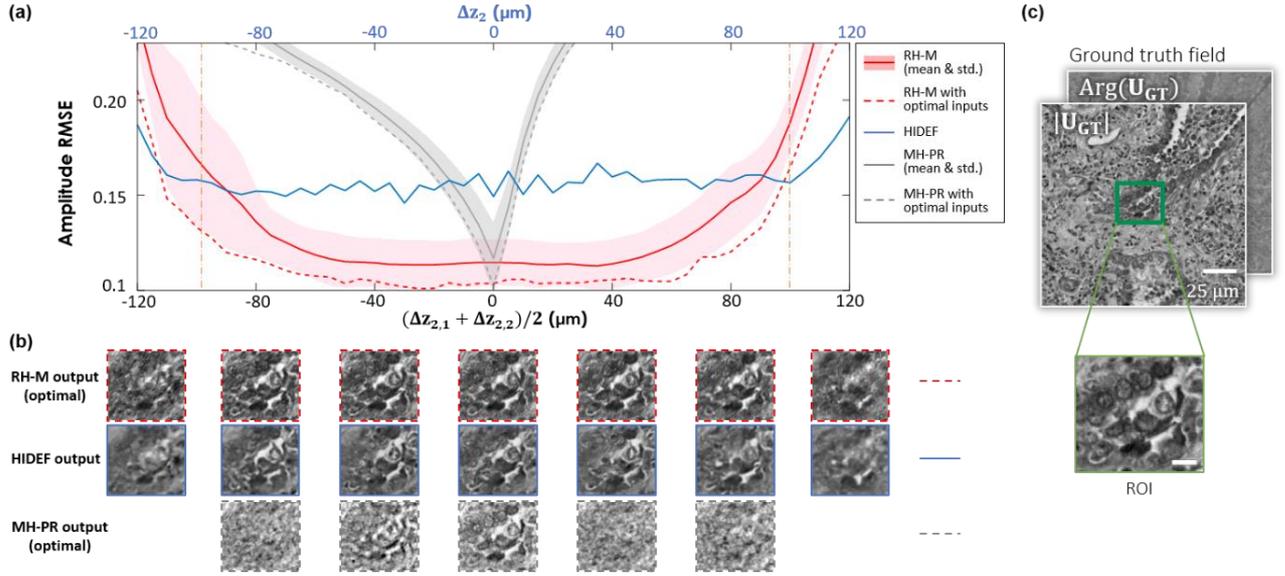

**Fig. 6** Extended DOF of RH-M. (a) The amplitude RMSE plot of the output complex fields generated by RH-M ($M=2$), HIDEF and MH-PR ($M=2$) as a function of the axial defocus distance; the ground truth field highlighted by the green rectangle in (c) is acquired using MH-PR with $M=8$ holograms. For RH-M ($M=2$) and MH-PR ($M=2$), mean and optimal RMSE values over different $\Delta z_{2,1}$, $\Delta z_{2,2}$ defocus combinations are shown. The orange dashed vertical lines show the axial training range for both HIDEF and RH-M. (b) Magnified ROIs of HIDEF outputs (blue) and optimal outputs of RH-M (red), and MH-PR (grey) at each defocus distance are shown below the plot. Outputs of MH-PR ($M=2$) with extremely low fidelity are omitted. (c) Ground truth complex field that is acquired using MH-PR with $M=8$ holograms. The ROI highlighted by the green rectangle is magnified. Scale bar: 10 μm.

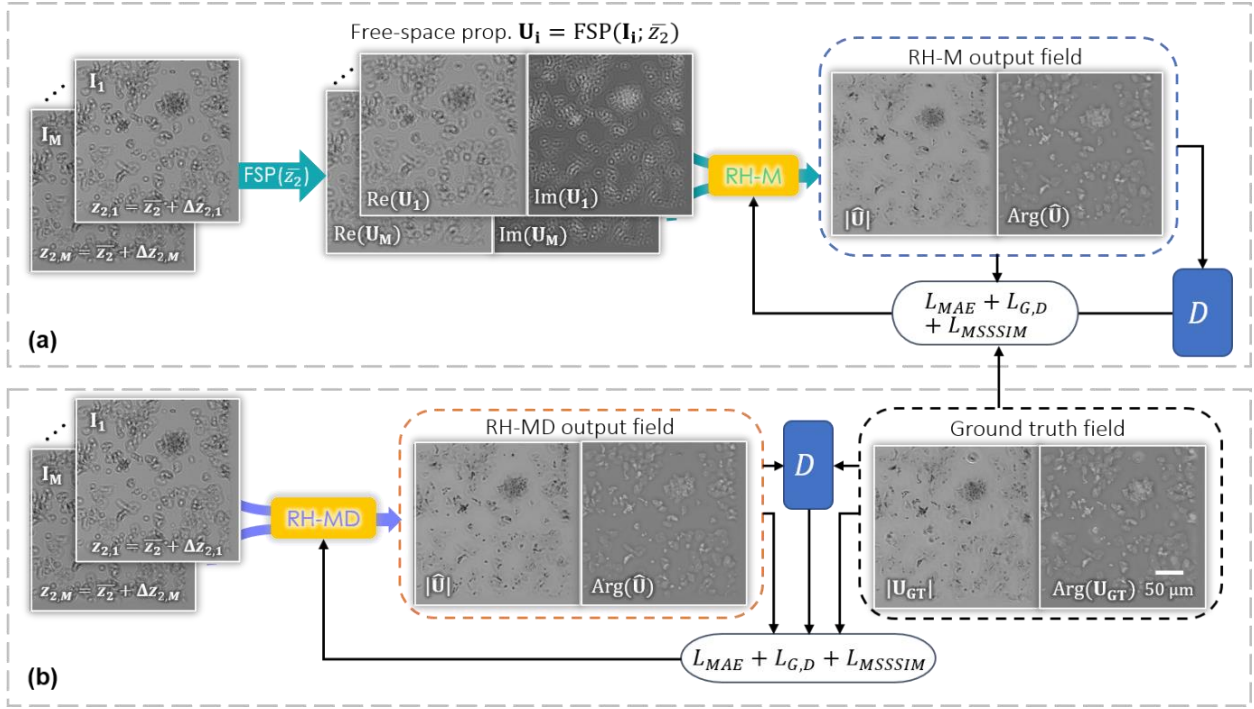

**Fig. 7** GAN framework used for training of RH-M and RH-MD. (a) GAN framework for training RH-M, which serves as the generator. *D* is the discriminator. (b) GAN framework for RH-MD. Generator and discriminator structures are depicted in Fig. 1 and are further detailed in the Materials and Methods section.

**Table 1** Inference time comparison of RH-M, RH-MD, HIDEF and MH-PR algorithms using Pap smear and lung tissue test datasets.

| Algorithm | # trainable parameters | Inference time (s/mm$^2$) |
|---|---|---|
| RH-M (M=2) | 14.1M | 0.162 |
| RH-M (M=3) | 14.1M | 0.200 |
| RH-MD (M=2) | 14.1M | 0.154 |
| MH-PR (M=2) | NA | 2.467 |
| MH-PR (M=3) | NA | 3.280 |
| MH-PR (M=8) | NA | 7.944 |
| HIDEF | 3.9M | 0.352 |

18